# Empirical Study of Sensor Observation Services Server Instances

Alain Tamayo, Pablo Viciano, Carlos Granell, Joaquín Huerta

Institute of New Imaging Technologies
Universitat Jaume I, Spain
Av. Vicent Sos Baynat, SN, 12071, Castellón de la Plana
{atamayo, pablo.viciano, carlos.granell, huerta}@uji.es

## Abstract

The number of Sensor Observation Service (SOS) instances available online has been increasing in the last few years. The SOS specification standardises interfaces and data formats for exchanging sensor-related information between information providers and consumers. SOS in conjunction with other specifications in the Sensor Web Enablement initiative, attempts to realise the *Sensor Web* vision, a worldwide system where sensor networks of any kind are interconnected. In this paper we present an empirical study of actual instances of servers implementing SOS. The study focuses mostly in which parts of the specification are more frequently included in real implementations, and how exchanged messages follows the structure defined by XML Schema files. Our findings can be of practical use when implementing servers and clients based on the SOS specification, as they can be optimized for common scenarios.

## 1 Introduction

The Sensor Web Enablement (SWE) initiative is a framework that specifies interfaces and metadata encodings to enable real time integration of heterogeneous sensor networks into the information infrastructure. It provides services and encodings to enable the creation of web-accessible sensor assets (Botts et al. 2008). It is an attempt to define the foundations for the Sensor Web vision, a worldwide system where sensor networks of any kind can be connected (van Zyl et al. 2009).

SWE includes specifications for service interfaces such as: Sensor Observations Service (SOS), a standard interface for requesting, filtering, and retrieving observations and sensor system information (OGC 2007); and Sensor Planning Service (SPS), a standard for requesting information about the capabilities of a sensor and for defining tasks over those sensors (OGC 2007a). It also includes encodings for the information exchanged between information providers and consumers. The main encodings are Observation and Measurement (O&M) (OGC 2007b), which defines standard models for encoding observations and measurements from a sensor; and the Sensor Model Language (SensorML) (OGC 2007c) defining standard models for describing sensor systems and processes. The format of the exchanged messages is defined using XML Schema, a language used to assess the validity of well-formed element and attribute information items contained in XML instance files (W3C 2004, 2004a).

The number of SOS server instances available online has been increasing in the last few years. Although these instances are based in the same implementation specification they frequently differ in subtle ways of representing information, for example, which subsets of the schemas they use, which protocols are used to request information, etc. These differences make *interoperability* a goal that is hard to achieve in practice.

In this paper, we present an empirical study of servers implementing SOS. The study focuses mostly in which parts of the specification are more frequently included in actual implementations, and how messages exchanged between clients and servers follow the structure defined by XML Schema files. The differences found between servers may shed some light to the cause of interoperability problems. The study may also show how different servers tend to group observations into offerings, or which spatial features are more often used to represent the offerings, just to mention two possible outcomes.

The remainder of the paper is structured as follows. Section 2 introduces the SOS specification and lists the server instances used later on subsequent sections. After this, Section 3 presents the result of the analysis of the information gathered from the sever instances. In this section we calculate the values of different metrics, such as number of invalid files, frequent validation errors, etc. Section 4 analyses which part of the schema files are used by the servers. Section 5 summarizes and discusses the results of the previous section. Lastly, we present the conclusions of our study.

## 2 Sensor Observation Services

The SOS specification provides a web service interface to retrieve sensor and observation data. The model used to represent the sensor observations defines the following concepts (OGC 2007, 2007b):
- *observation*: act of observing a property or phenomenon, with the goal of producing an estimate of the value of the property
- *feature of interest*: feature representing the real world object which is the observation target
- *observed property*: phenomenon for which a value is measured or estimated
- *procedure*: process used to produce the result. It is typically linked to a sensor or system of sensors
- *observation offering*: logical grouping of observations offered by a service that are related in some way

The operations of the SOS specification are divided into three profiles (OGC 2007):
- *Core profile*: mandatory operations for any SOS server instance:
    o *GetCapabilities*: It retrieves metadata information about the service.
    o *DescribeSensor*: It retrieves information about a given procedure.
    o *GetObservation*: It retrieves a set of observations that can be filtered by a time instant or interval, location, etc.
- *Transactional profile*: optional operations for data producers to interact with the server:
    o *RegisterSensor*: It allows new sensors to be inserted.
    o *InsertObservation*: It allows new observations to be inserted.
- *Enhanced profile*: optional profile including a richer set of operations to interact with the server. For example:
    o *GetFeatureOfInterest*: Returns the geometry describing a feature of interest.

    o *GetResult*: It allows clients to reduce the transfer of redundant information related with sensor data when working with the same set of sensors.

 The information about sensors and observations retrieved from the servers is usually encoded using SensorML (OGC 2007c) and O&M (OGC 2007b). Nevertheless, the specification allows data producers to encode data in their own favourite formats. SOS also depends for its implementation on other specifications such as Geography Markup Language (GML) (OGC 2004), OGC Web Services Common (OGC 2007d) and Filter encoding specification (OGC 2005). All of these dependencies are shown in Figure 1.

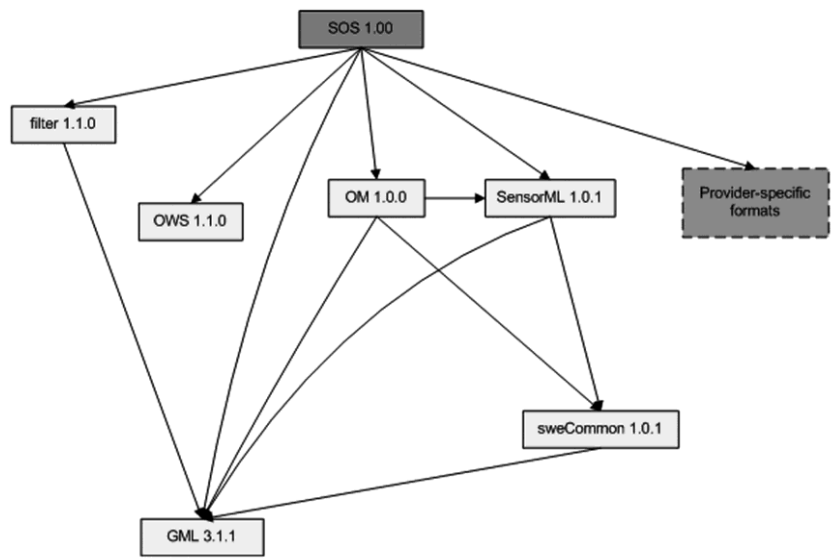

**Fig 1:** Dependencies of SOS from other specifications

## 2.1   SOS server instances

In order to realize our study we gathered information from a set of SOS server instances freely available on the Internet. The URLs of these servers are listed in Table 1. These servers where located using web services cata-logs such as the OWS Search Engine[1] and IONIC RedSpider Catalog Client[2], and using general-purpose search engines such as Google and Yahoo!. The table only shows the servers claiming to support version 1.0.0 of the standard.

**Table 1:** List of SOS server instances

| | Server URL |
|---|---|
| 1 | http://152.20.240.19/cgi-bin/oos/oostethys_sos.cgi |
| 2 | http://204.115.180.244/server.php |
| 3 | http://81.29.75.200:8080/oscar/sos |
| 4 | http://ak.aoos.org/ows/sos.php |
| 5 | http://bdesgraph.brgm.fr/swe-kit-service-ades-1.0.0/REST/sos |
| 6 | http://ccip.lat-lon.de/ccip-sos/services |
| 7 | http://compsdev1.marine.usf.edu/cgi-bin/sos/v1.0/oostethys_sos.cgi |
| 8 | http://coolcomms.mote.org/cgi-bin/sos/oostethys_sos.cgi |
| 9 | http://data.stccmop.org/ws/util/sos.py |
| 10 | http://devgeo.cciw.ca/cgi-bin/mapserv/sostest |
| 11 | http://elcano.dlsi.uji.es:8080/SOS_MCLIMATIC/sos |
| 12 | http://esonet.epsevg.upc.es:8080/oostethys/sos |
| 13 | http://gcoos.disl.org/cgi-bin/oostethys_sos.cgi |
| 14 | http://gcoos.rsmas.miami.edu/dp/sos_server.php |
| 15 | http://gcoos.rsmas.miami.edu/sos_server.php |
| 16 | http://gis.inescporto.pt/oostethys/sos |
| 17 | http://giv-sos.uni-muenster.de:8080/52nSOSv3/sos |
| 18 | http://habu.apl.washington.edu/cgi-bin/xan_oostethys_sos.cgi |
| 19 | http://lighthouse.tamucc.edu/sos/oostethys_sos.cgi |
| 20 | http://mmisw.org/oostethys/sos |
| 21 | http://nautilus.baruch.sc.edu/cgi-bin/sos/oostethys_sos.cgi |
| 22 | http://neptune.baruch.sc.edu/cgi-bin/oostethys_sos.cgi |
| 23 | http://oos.soest.hawaii.edu/oostethys/sos |
| 24 | http://opendap.co-ops.nos.noaa.gov/ioos-dif-sos/SOS |
| 25 | http://rtmm2.nsstc.nasa.gov/SOS/footprint |
| 26 | http://rtmm2.nsstc.nasa.gov/SOS/nadir |
| 27 | http://sccoos-obs0.ucsd.edu/sos/server.php |
| 28 | http://sdf.ndbc.noaa.gov/sos/server.php |
| 29 | http://sensor.compusult.net:8080/SOSWEB/GetCapabilitiesGFM |
| 30 | http://sensorweb.cse.unt.edu:8080/teo/sos |
| 31 | http://sensorweb.dlz-it-bvbs.bund.de/PegelOnlineSOS/sos |
| 32 | http://sos-ws.tamu.edu/tethys/tabs |
| 33 | http://swe.brgm.fr/constellation-envision/WS/sos-discovery |
| 34 | http://vast.uah.edu/ows-dev/dopplerSos |

---

[1] http://ows-search-engine.appspot.com/index
[2] http://dev.ionicsoft.com:8082/ows4catalog/elements/sos.jsp

| | |
|---|---|
| 35 | http://vast.uah.edu/ows-dev/tle |
| 36 | http://vast.uah.edu/vast/nadir |
| 37 | http://vast.uah.edu:8080/ows-dev/footprint |
| 38 | http://vastserver.nsstc.uah.edu/vast/adcp |
| 39 | http://vastserver.nsstc.uah.edu/vast/airdas |
| 40 | http://vastserver.nsstc.uah.edu/vast/weather |
| 41 | http://v-swe.uni-muenster.de:8080/WeatherSOS/sos |
| 42 | http://weather.lumcon.edu/sos/server.asp |
| 43 | http://webgis2.como.polimi.it:8080/52nSOSv3/sos |
| 44 | http://wron.net.au/BOM_SOS/sos |
| 45 | http://wron.net.au/CSIRO_SOS/sos |
| 46 | http://ws.sensordatabus.org/Ows/Swe.svc/ |
| 47 | http://www.cengoos.org/cgi-bin/oostethys_sos.cgi |
| 48 | http://www.csiro.au/sensorweb/BOM_SOS/sos |
| 49 | http://www.csiro.au/sensorweb/CSIRO_SOS/sos |
| 50 | http://www.csiro.au/sensorweb/DPIW_SOS/sos |
| 51 | http://www.gomoos.org/cgi-bin/sos/V1.0/oostethys_sos.cgi |
| 52 | http://www.mmisw.org:9600/oostethys/sos |
| 53 | http://www.pegelonline.wsv.de/webservices/gis/sos |
| 54 | http://www.wavcis.lsu.edu/SOS/server.asp |
| 55 | http://www.weatherflow.com/sos/sos.pl |
| 56 | http://www3.gomoos.org:8080/oostethys/sos |

Starting from these servers we retrieved a sample set of XML instance files including service metadata, and sensors and observations information. These instance files were then analysed mainly regarding to schema validity and used features. The results from this analysis are shown extensively in Section 3.

### 2.2 Limitations of the Study

This study presents some limitations. First, it is impossible to retrieve all of the information published on the servers. We tried to overcome the effects of this limitation by making the sample dataset as large as possible and, in cases where several alternatives exists for making a request, we retrieved at least one instance file from each alternative. Second, only responses from the core profile operations were considered. This is because most servers do not implement the rest of the operations (see Section 3.1.2). Third, we did not test server instances for full compliance to the SOS specification; we only deal with the information contained in the XML instance files and XML schema files. Last, we analysed server instances without considering the server product used to deploy the instance. This is because for several instances we were not able to determine which product was used, and in some cases a handcrafted servers have been developed for specific problems.

## 2.3 Dataset Description

Details about the information contained in the sample dataset are presented in Table 2. The table includes the following information for the responses of the considered operations:
- *Number of files (NF):* Number of files retrieved for the operation.
- *Number of objects described (NO)*: Depending on the operations these objects are *observations offerings*, in the case of the *GetCapabilities* operation; *sensor systems*, in the case of *DescribeSensor*; and *observations*, in the case of *GetObservation*.

**Table 2:** Dataset description

| Operation | NF | NO |
|---|---|---|
| GetCapabilities | 56 | 7190 |
| DescribeSensor | 6719 | 6719 |
| GetObservation | 204 | 3990656 |
| Total | 6979 | 4004565 |

## 3 Results

In this section we present the results of computing the sample dataset according to the following metrics:
- *Number of Invalid Files:* Number of files invalid according to the schema files.
- *Most frequent validation errors:* List with most frequent errors found during validation, including an error description and the number of occurrences of each error.
- *Used Features:* The features presented depend on the analysed operation. For example, while analysing capabilities files we considered supported operations or filters and response formats. While analysing observation files, we considered, for example, which observation type is most frequently used to encode the information gathered by sensors.
- *Parts of the schemas that are actually used*: Schema files defining the message structures for SOS are large and complex, moreover, SOS schema files depend on schema files included on other specifications as well. For these reasons actual implementations only use a subset of these schemas.

We present the results of applying the first three metrics divided by operation. Then, in a different section we analyse the part of schemas that are actually used.

### 3.1 Capabilities files

The capabilities file of a server contains all of the information needed to access the data it contains. In the case of SOS servers, this file contains available observation offerings, supported operations and filters, etc.

#### 3.1.1 Instances validation

The first important fact extracted from the sample dataset is that 34 out of 56 (60.7%) capabilities files are *invalid* according to the schemas defining their structure. Table 3 shows the most frequent errors found in the instance files.

**Table 3:** Most frequent validation errors for capabilities files

|   | Error code | Description | Number of Occurrences |
|---|---|---|---|
| 1 | cvc-complex-type.2.4.a | Invalid content was found starting with element [element name]. One of {valid element list} is expected | 2,754 |
| 2 | cvc-complex-type.2.2 | Element must have no element [children], and the value must be valid | 978 |
| 3 | cvc-datatype-valid.1.2.3 | [value] is not a valid value of union type | 960 |
| 4 | cvc-attribute.3 | The value of attribute on element is not valid with respect to its type | 468 |
| 5 | cvc-datatype-valid.1.2.1 | [value] is not a valid value of union type | 379 |
| 6 | cvc-id.2 | There are multiple occurrences of ID value | 107 |

The most frequent error found was the use of a different name for an element that the one specified in the schemas. For example, this was the case for element *sos:Time*, which specifies the time instant or period for the observations within an offering. The element name was changed to

*sos:eventTime* in some of the servers, maybe because that was the name in previous versions of the specification. The second most frequent error was elements with invalid content (errors 2, 3, 4 and 5). Common mistakes were time values with incorrect format, or offering ID values containing whitespaces or colons.

Despite of the large number of errors found, most of them did not prevent the files from being correctly parsed, although they supposed and extra amount of work while implementing the parsers. At the end only 2 of the 56 files contained serious errors, which make parsing their content impossible for us.

### 3.1.2 Supported Operations

The capabilities files also indicate which operations are supported by the servers, including information about how to access them and which values are allowed as parameters. Table 4 shows which and how frequently the different operations are supported.

**Table 4:** Operations supported for the server instances

|    | Operation Name | Profile | GET Support | POST Support |
|----|----------------|---------|-------------|--------------|
| 1  | GetCapabilities | Core | 56 | 54 |
| 2  | DescribeSensor | Core | 33 | 45 |
| 3  | GetObservation | Core | 42 | 54 |
| 4  | RegisterSensor | Transactional | 0 | 2 |
| 5  | InsertObservation | Transactional | 1 | 2 |
| 6  | GetFeatureOfInterest | Enhanced | 0 | 12 |
| 7  | GetObservationById | Enhanced | 0 | 10 |
| 8  | GetResult | Enhanced | 0 | 1 |
| 9  | GetFeatureOfInterestTime | Enhanced | 0 | 0 |
| 10 | DescribeFeatureType | Enhanced | 0 | 0 |
| 11 | DescribeObservationType | Enhanced | 0 | 0 |
| 12 | DescribeResultModel | Enhanced | 0 | 0 |

The results, also depicted in Figure 2, show that all of the servers implement the *GetCapabilities* request using HTTP GET as required by the SOS implementation specification. Apart from that, most of them also implement the operation using HTTP POST. Most complex requests such as *GetObservation* are implemented easier using HTTP POST than using HTTP GET, as the SOS specification does not define KVP encodings for this operation.

The core profile is mandatory for every server but 10 of the 56 servers do not implement the *DescribeSensor* request, or at least they do not include it in the capabilities file. Operations for the transactional and enhanced profile are implemented by a few server instances and some of them are not implemented at all.

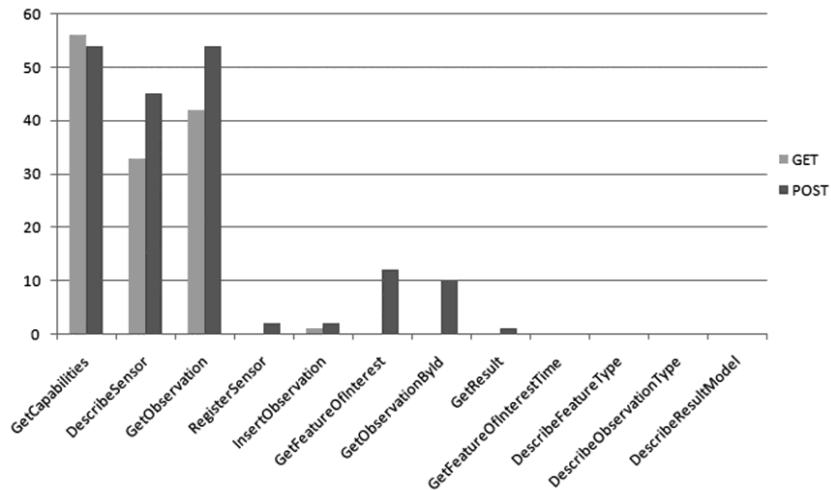

**Fig 2:** Support of SOS operations in actual server instances

### *3.1.3 Supported Filters*

The number of potential observations published on a server can be very large. For this reason, filters are used to request just the observations in which we are interested. Filters for SOS fall into four categories: spatial, temporal, scalar, and identifier filters. Only 16 of the 56 (28.5%) capabilities files include information about the supported filters. These filters are detailed in Table 5. For each filter category the supported operands and operators are shown, as well as how frequently they have been used.

The most implemented filters are *BBOX* and *TM_During* that allow to restrict the location of the observations to a given bounding box or to a given time period respectively. *Id filters* are also frequently implemented. They allow information be filtered by specifying the ID of entities related with the request. Even though some servers do not include the filter capabilities section, most of them allow observations to be also filtered using a bounding box or a time interval.

**Table 5:** Support of filters

| Filter Category | | | Number of Appearances |
|---|---|---|---|
| Spatial Filters | *Operands* | gml:Envelope | 16 |
| | | gmlPolygon | 11 |
| | | gml:Point | 11 |
| | | gml:LineString | 11 |
| | *Operators* | BBOX | 15 |
| | | Contains | 11 |
| | | Intersects | 11 |
| | | Overlaps | 11 |
| | | Equals | 1 |
| | | Disjoint | 1 |
| | | Touches | 1 |
| | | Within | 1 |
| | | Crosses | 1 |
| | | DWithin | 1 |
| | | Beyond | 1 |
| Temporal Filters | *Operands* | gml:TimeInstant | 16 |
| | | gml:TimePeriod | 16 |
| | *Operators* | TM_During | 15 |
| | | TM_Equals | 14 |
| | | TM_After | 14 |
| | | TM_Before | 14 |
| | | TM_Begins | 1 |
| | | TM_Ends | 1 |
| Scalar Filters | *Operators* | Between | 14 |
| | | EqualTo | 13 |
| | | NotEqualTo | 13 |
| | | LessThan | 13 |
| | | LessThanEqualTo | 13 |
| | | GreaterThan | 13 |
| | | GreaterThanEqualTo | 13 |
| | | Like | 12 |
| | | NullCheck | 1 |
| Id Filters | | eID | 16 |
| | | fID | 15 |

### 3.1.4 Supported Response Formats

Observations published on different server instances are encoded using several different formats. These formats and the number of offerings that represent information with them are presented in Table 6. The most supported format to represent observations is O&M 1.0.0, which is the default format specified by SOS. A deeper discussion about this format is presented in Section 3.3

**Table 6:** Formats supported to represent observation information

| Format | Number |
|---|---|
| text/xml; subtype="om/1.0.0" | 5110 |
| text/xml;schema="ioos/0.6.1" | 2064 |
| text/csv | 664 |
| application/vnd.google-earth.kml+xml | 664 |
| text/tab-separated-values | 664 |
| application/zip | 110 |
| text/xml | 4 |
| application/com-binary-base64 | 1 |
| application/com-tml | 1 |

### 3.1.5 Offerings Information

Observation offerings contain information about a set of related sensor observations. The SOS specification does not say how observations, procedures or observed properties should be grouped into offerings. For this reason, it would be very interesting to know how this grouping is realised in actual implementations. Regarding observation offerings we computed the followed metrics:

- *Number of offerings per server (OpS)*: How many offerings are usually published on a server
- *Number of procedures per server (PpS)*: How many sensor or sensor systems are published on a server
- *Number of observed properties per server (OPpS)*: How many observed properties are usually published on a server
- *Number of offering as points*: An interesting peculiarity observed during the experiments is that location of most offerings is a point, instead of a bounding box.

The result of computing the first three metric values is shown in Figure 3. The figure shows the values grouped into 6 categories. The number of offerings per server ranges from 1 to 1772. 48% of the servers contain 1-4 offerings, and 63 % contain 16 or less. This indicates than servers tend to group observations in a few offerings. Similarly, the number of procedures per server ranges from 1 to 1957. Although in a lesser degree than the case of offerings the number of servers with a large amount of procedures per server is always lower than the number of server with a small number of procedures. The number of observed properties per server ranges from 1 to 114. This number behaves much like the previous ones having 65% of the server instance with less than 16 observed properties advertised.

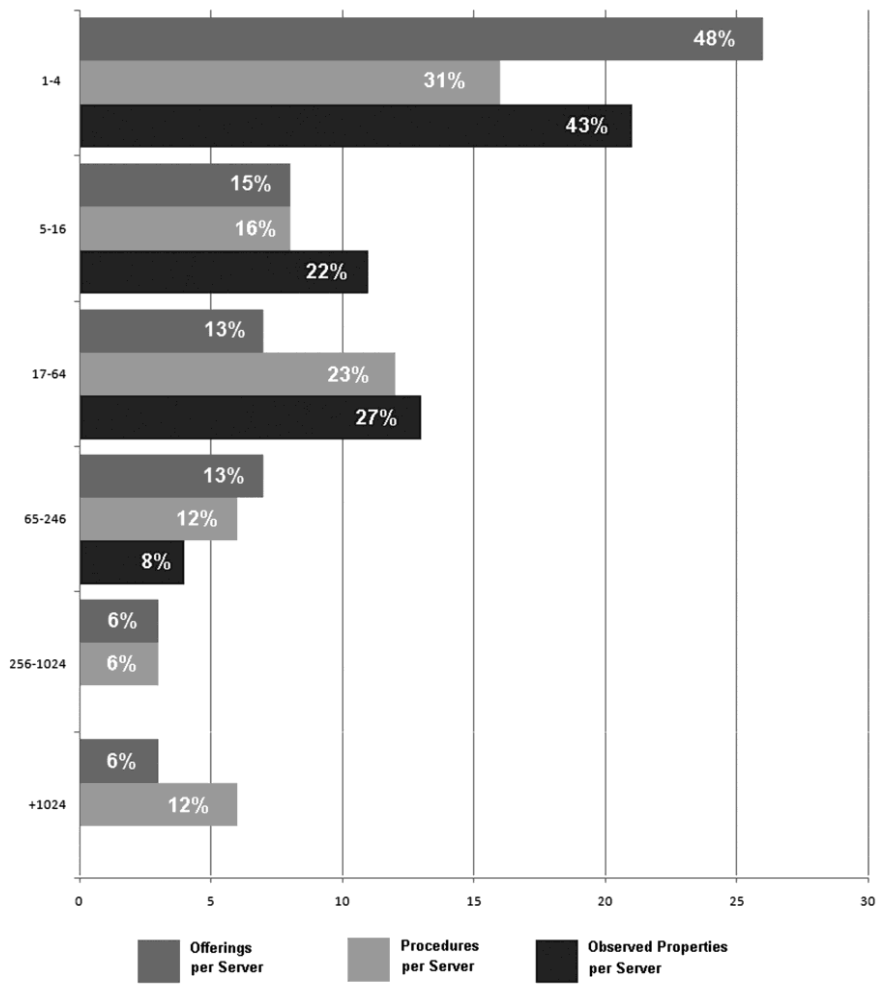

**Fig 3:** Number of servers classified for the number of offerings, procedures and observed properties that they contain

A last interesting phenomenon found here is the number of observation offerings which are restricted to a point in the space. Each offering has a property named *boundedBy* defining a bounding box where the observations grouped in the offering are located. In 6575 offerings in the sample data set this bounding box was indeed a point, representing the 95.7% of the total number of offerings. This clearly indicates that the first criteria followed to group observations into offerings is the sensors location, which in most of the cases is a single point on the Earth. Figure 4 shows as an example a set of offerings located in North America represented in Google Earth. In the figure, *placemarks* represent *point offerings* and bounding boxes represent other offerings.

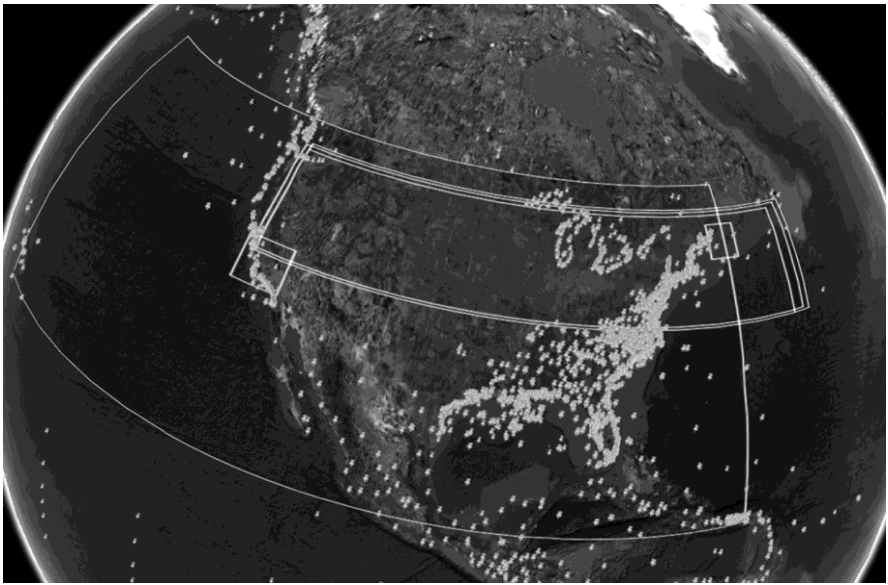

**Fig 4**: Observation offerings in North America

### 3.2 Procedure description files

The 56 servers considered in this study mention in their capabilities files 12,222 procedures. From this number we were able to retrieve the description files of 6719 of them (54.9%). All of these files were encoded in the sensorML format.

### 3.2.1 Instances validation

The validation of the sensorML files gave as result that 1896 of the files were invalid according to the XML schemas files defining the structure of these documents. The value represents the 28.2% of the overall number of files. The most frequent errors found are presented in Table 7.

The first error type occurred frequently because required elements where omitted or elements not defined in the schemas where introduced in the wrong place. Errors 2, 4 and 5, similarly to the case of capabilities files refers to incorrect formatted values: identifiers including whitespaces or colons, incorrect time values, or values just being left empty. The most serious errors were those of type 2. In these cases, wrong use of namespaces, or not specifying the version of the schemas used, made impossible to process the documents at all.

**Table 7:** Most frequent validation errors for sensor description files

|   | Error code | Error Description | Number of Occurrences |
|---|---|---|---|
| 1 | cvc-complex-type.2.4.a | Invalid content was found starting with element [element name]. One of {valid element list} is expected | 1778 |
| 2 | cvc-attribute.3 | The value of attribute on element is not valid with respect to its type | 556 |
| 3 | cvc-elt.1.a | Cannot find the declaration of element [element name] | 500 |
| 4 | cvc-datatype-valid.1.2.1 | [value] is not a valid value of union type | 300 |
| 5 | cvc-pattern-valid | Value is not facet-valid with respect to pattern for type | 256 |

### 3.2.1 Procedure description types

The sensorML specification models sensor systems as a collection of physical and non-physical processes. Physical processes are those where information regarding their positions and interfaces may be relevant. Examples of these processes are detectors, actuators, and sensor systems. Non-physical or "pure" processes according to the specification "*can be treated as merely mathematical operations*" (OGC 2007c). These categories are further subdivided as shown next:

- *Physical processes*

- o *Component*: Any physical process that cannot be subdivided into smaller subprocesses.
- o *System*: It may group several physical or non-physical processes.
- *Non-physical processes*
    - o *Process Model*: Defines an atomic pure process which is used to form process chains
    - o *Process Chains*: Collection of executable processes in a sequential manner to obtain a desired result.

From 6219 processed sensorML files, 6215 described *Systems* (99.9%), and 4 of them *ProcessChains*. This indicates that the usual is to describe sensor systems that have a location in space and measure an observed property for a period of time.

### 3.2.1 Specifying location

An important piece of information about the procedure is its *location*. Unfortunately for programmers, location can be specified in different parts of the procedure description file (sensorML file). In the sample dataset we have found this information located in at least three different places and using different names to identify coordinates:

- Under the *location* tag in the description of a *System* as a point:

```
<SensorML xmlns="http://www.opengis.net/sensorML/1.0.1"
  version="1.0.1" [Other attributtes]>
  <member>
    <System gml:id=[System ID]>
      ...
      <location>
        <gml:Point srsName=[SRS Name]>
          <gml:coordinates>39.99 -0.068 0</gml:coordinates>
        </gml:Point>
      </location>
      ...
    </System>
  </member>
</SensorML>
```

- Under the *position* tag in the description of a *System* as a vector with named elements:

```
<SensorML xmlns="http://www.opengis.net/sensorML/1.0.1"
  version="1.0.1" [Other attributtes]>
  <member>
    <System gml:id=[System ID]>
      ...
      <sml:position name=[name]>
        <swe:Position referenceFrame=[SRS name]>
```

```
            <swe:location>
              <swe:Vector>
                <swe:coordinate name="x">
                  <swe:Quantity>
                    <swe:value>-0.068</swe:value>
                  </swe:Quantity>
                </swe:coordinate>
                <swe:coordinate name="y">
                  <swe:Quantity>
                    <swe:value>39.99</swe:value>
                  </swe:Quantity>
                </swe:coordinate>
                <swe:coordinate name="z">
                  <swe:Quantity>
                    <swe:value>0</swe:value>
                  </swe:Quantity>
                </swe:coordinate>
              </swe:Vector>
            </swe:location>
          </swe:Position>
        </sml:position>
        ...
      </System>
  </member>
</SensorML>
```

- Under the *positions* tag in the description of a *System* as a list of positions:

```
<SensorML xmlns="http://www.opengis.net/sensorML/1.0.1"
  version="1.0.1" [Other attributtes]>
  <member>
    <System gml:id=[System ID]>
      ...
      <sml:positions>
        <sml:PositionList>
          <sml:position name=[name position 1]>
            [Position data]
          </sml:position>
          <sml:position name=[name position 2]>
            [Position data]
          </sml:position>
          ...
        </sml:PositionList>
      </sml:positions>
      ...
    </System>
  </member>
</SensorML>
```

In the first case reading the coordinates values is straightforward, the values are grouped together into a *gml:Point* object. In the second one, several tags must be parsed to reach the coordinates; a problematic issue at this point is that different names are used by servers to refer to the coordinate values. For example *longitude* was also named *x* or *easting*; *latitude* was also named *y* or *northing*; and *altitude* was also named *z*. The contents and attributes of the tags involved are also slightly different, some servers includes unit of measurements, some include the axis they refer to, etc. The third case is a generalization of the second one, where positions are included in a list, allowing more than one to be specified. None of the analysed files included more than one position for a sensor or sensor system.

## 3.3 Observation files

To analyse *GetObservation* responses, 1.7 GB of observation data was retrieved from the server instances. All of the retrieved files follow the format specified by O&M 1.0.0 encoding specification. As shown in Table 6 this is most widely used format and is the default for encoding observations in SOS 1.0.0.

### 3.3.1 Instances validation

Validation of observation files was much more difficult than expected. The validation process failed repeatedly to process correctly large files (> 10MB) and did not allow the validation of files containing *measurements* alleging that schema files were incorrect. *Measurements* are specialized observations where the observation value is described using a numeric amount with a scale or using a scalar reference system (OGC 2007b). Large files were only a few, so the first limitation was not a great problem but files containing measurements were about half of the whole observation files. Although we were able to parse correctly all of the observations, we were only able to apply the validation process to 62 files (31.3%). From these 62 files, 56 were reported to be invalid (90%). Details about the errors found are shown in Table 8.

**Table 8:** Most frequent validation errors for sensor description files

|   | Error code | Error Description | Number of Occurrences |
|---|---|---|---|
| 1 | cvc-attribute.3 | The value of attribute on element is not valid with respect to its type | 206 |
| 2 | cvc-complex-type.2.4.a | Invalid content was found starting with element [element name]. One of {valid | 189 |

| | | element list} is expected | |
|---|---|---|---|
| 3 | cvc-datatype-valid.1.2.1 | [value] is not a valid value of union type | 121 |

The validation errors for observation files are similar to those for capabilities and sensor description files. Values with wrong formats, and wrong named or misplaced elements made up all of the errors found in the instance files.

### 3.3.2 Observation Types

According to the O&M 1.0.0 encoding specification observation types are organized as shown in Figure 5. The base type for all observations is *ObservationType*, which inherits form *AbstractFeatureType* located in GML schemas. Starting from *ObservationType* a set of specializations is defined based on the type of the results contained in the observations. Additionally, information providers can derive their own observation data types from the different types in the figure.

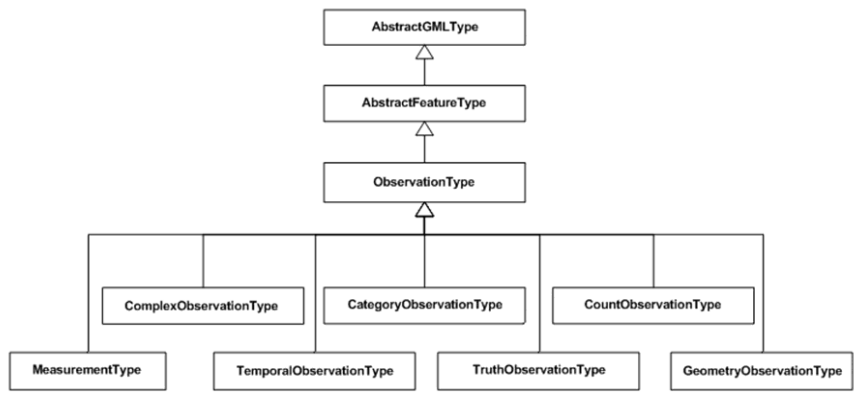

**Fig 5**: Hierarchy of observation types

From 3,990,656 observation values processed in the dataset, 56.3 % (2,246,639) of the values were *Observation* elements (instances of *ObservationType)*, and 43.7 % (1,744,017) were instances of *Measurement* elements (instances of *MeasurementType*). Values corresponding to none of the other types were found in the sample dataset. Despite of the fact that the number of measurement values was lower than the number of observations, the amount of disk space needed to contain these values was about 7 times larger than the space occupied by the observations (1533 MB against 213 MB). This difference in size seems to be the cause why most implementations choose not to use observation specialization types. Although the lack in the O&M specification of well-defined semantic models might influence this decision as well (Probst 2008, Kuhn 2009).

## 3.4 Subset of XML Schemas used

The last piece of information we extracted from the sample dataset is the subset of the XML Schemas that is actually used in the server instances. The number of schema files associated to the SOS specification is huge. If we follow all of the dependencies from the main schema files of the specification we obtain a set of 87 files. If we additionally consider the observations specialization schemas (containing the definition of *MeasurementType*) and their own dependencies this number grows to 93. The size of schemas bring as a consequence that server instances only provide support for a subset of them.

Next, we calculate from the sample dataset which part of the schemas is used and which part is not used at all. To calculate this information we inspect the information contained on the instance files to determine which schema components are directly used in the files (*initial set*). After doing this, we determine which other schema components are used to define the initial set. The algorithm used is similar to the one included in the GML subsetting profile tool, a tool used to extract subsets of the GML schemas (OGC 2004). We present the results in two steps. Firstly, we detail the subset of the GML schemas that is actually used. And secondly, a similar analysis with the overall results for the SOS specification is presented.

### 3.4.1 GML

GML constitutes more than 50% of the overall number of global schema components (types, elements, model groups) comprising the SOS schemas. It is used to model geographic features embedded into the instance files, and its components are extended or composed into new components of the SOS specification. As shown in Figure 1, most of the specifications relevant to our study depend to a large extent on GML.

Table 9 shows a comparison between the number of components in the original GML schemas for version 3.1.1 (*original files*) and the subset of the schemas that is referenced directly, or used in the definition of others components referenced directly in the sample dataset (*profile*).

The results are divided by component type: complex types (#CT), simple types (#ST), global elements (#EL), global attributes (#AT), model groups (#MG) and attribute groups (#AG). It turned out that only 16.3% of the components were actually used. All of the components contained in the following files were not used at all: *coverage.xsd, dataQuality.xsd, defaultStyle.xsd, direction.xsd, dynamicFeature.xsd, geometricComplexes.xsd, geometricPrimitives.xsd, grids.xsd, measures.xsd, temporalreferenceSystems.xsd, temporalTopology.xsd, topology.xsd, and valueObjects.xsd*.

**Table 9:** Comparison between overall number of components and number of components actually referenced in GML

|       | Original files | Profile |
|-------|----------------|---------|
| #CT   | 394            | 60      |
| #ST   | 64             | 15      |
| #EL   | 485            | 74      |
| #AT   | 15             | 9       |
| #MG   | 12             | 2       |
| #AG   | 35             | 4       |
| Total | 1005           | 164     |

### 3.4.2 SOS

As mentioned before, the full SOS schemas are comprised by 93 files, distributed by specification as presented in Table 10. This full set is calculated starting from the SOS *"main schemas"* and following the references specified with *include* and *import* tags. For example, a typical practice when accessing a component in the GML schemas is to import the whole schemas through the file *gml.xsd*. This way all of the GML schemas become referenced even when most of them are never used.

**Table 10:** Distribution of SOS 1.0.0 schema files by specification

| Specification | Version | Number of files |
|---------------|---------|-----------------|
| SOS           | 1.0.0   | 16              |
| GML           | 3.1.1   | 32              |
| SensorML      | 1.0.1   | 5               |
| OM            | 1.0.0   | 3               |
| SWE Common    | 1.0.1   | 11              |
| Sampling      | 1.0.0   | 5               |
| OWS           | 1.1.0   | 14              |
| Filter        | 1.1.0   | 4               |
| Others        |         | 3               |

Table 11 shows a comparison between overall number of components in the full SOS schemas (*original files*) and the subset of the components that is really needed as explained before in the case of GML (*profile*). The results are also displayed in Figure 6.

**Table 11:** Comparison between overall number of components and number of components actually used in the SOS full schema set

| Metric | Original files | Profile |
|---|---|---|
| #CT | 772 | 266 |
| #ST | 119 | 61 |
| #EL | 745 | 201 |
| #AT | 39 | 3 |
| #MG | 28 | 16 |
| #AG | 40 | 8 |
| Total | 1743 | 515 |

Only 29.5% of the components in the full schema set are actually used in the sample dataset.

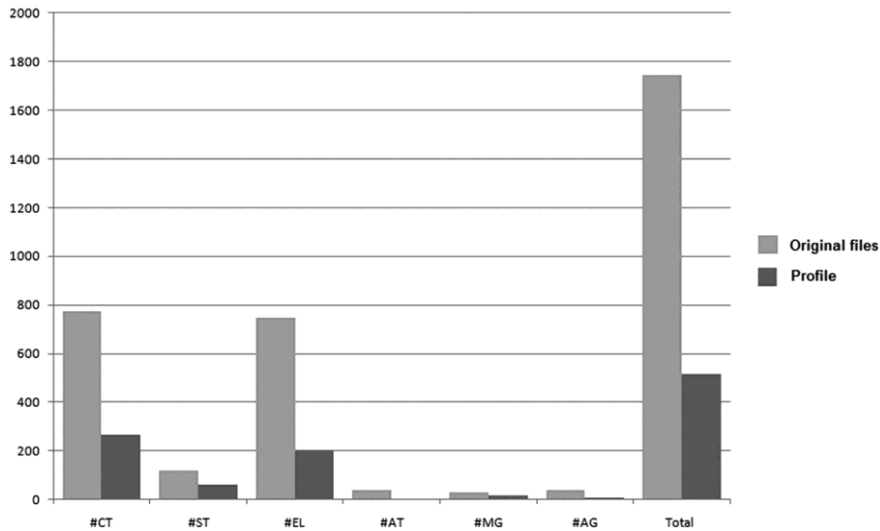

**Fig 6**: Overall number of schema components vs actually used components in SOS.

## 4 Discussion

As the amount of information extracted from the sample dataset is large we present a summary of our findings:

1. The number of invalid instances files is high: 29% (1986 out of 6837),
2. Most of the validation errors found are not serious enough to prevent correct parsing in many cases,
3. Some servers do not implement all of the mandatory operations in the core profile,
4. Most servers do not advertised filtering capabilities,
5. Most servers use O&M to encode observations,
6. Most servers group observations into a small number of offerings, and they usually contain information about a small number of procedures and observation properties,
7. Offerings location is frequently a point in space indicating that the first criteria to distribute observations into offerings is the sensors location,
8. Most procedure descriptions refer to *Systems*,
9. All of the observations in the sample dataset belong to only two types: *observations* and *measurements*,
10. The size on disk needed to represent *measurements* is a lot higher than to represent the same information as basic *observations*.
11. Most servers only support operations from the core profile,
12. Procedure location is specified in at least three different parts of the sensorML documents, and sometimes coordinates are referred to under different names. This problem could be solved by allowing only one of the three choices. If multiple locations can be specified for a procedure the more general solution would be the most appropriate, although we did not find any instance with more than one location in the sample dataset.
13. Only 29.5 % of the full schema set for SOS is used by the sample dataset

The first four points are closely related to interoperability. The presence of invalid files increases the chance of parsing errors in client-side applications. The fact that most errors are easy to overcome if writing the parsers manually, do not deny the fact that may limit the applicability of XML data binding code generators if they are strict regarding schemas validity. Not supporting mandatory operations may also lead a client to fail if they request these operations to the server. Not advertising filtering capabilities simply prevent the clients to effectively filter the observations, unless they know beforehand how the server works.

Next six points (5-10) provides useful insight for optimizing server and client implementations. Knowing which formats, offering grouping strategies, and types of sensor and observation representations are more commonly used could be utilised to optimize implementations to these scenarios. Even more, they could indicate which features are most likely to stay in future versions of the specification. Point 10 is specially revealing if large amounts of information are being handled. In this case using measurements are not the right choice for encoding information.

Last three points, in our opinion, reflect the complexity of the SOS specification. The number of operations in the specification is high if compared with others OGC specifications. In addition, the complexity of the formats that most be supported such as SensorML, O&M, SWE common, and GML, makes the implementation of the core profile itself a complex task. The example of how location is specified for procedures shows that even getting a simple piece of information can be a difficult thing to do. The last point could be the result of two options: the schemas are too complex to be implemented in its entirety or most of the information included or referenced by the schemas is not needed in real scenarios. In our opinion both options are true to a certain degree. Schemas are complex enough to make almost impossible to fully implement them manually. This complexity also makes code generation based on them tricky, as they use schema features that are not supported by some generators. In addition, some of schemas contain validation errors. Regarding if all of the information included in the schemas are really needed, they have been designed to be useful in as many scenarios as possible. Even if the design process starts with a very well defined use cases, how real users are going to utilise them is not easy to predict.

## 5 Conclusions

In this paper we have presented an empirical study of actual instances of servers implementing SOS. The study have focused mostly in which parts of the specification are more frequently included in real implementations, and how exchanged messages follows the structure defined by XML Schema files. Several interesting outcomes have been obtained such as the main criteria to group observations into offerings, the small subset of the schemas that are actually used, the large number of files that are invalid according to the schemas, etc.

All of these findings must be taken with care because the study have presented several limitations such as the impossibility to retrieve all of the information published on servers or only the responses from the core profile operations were considered. Nevertheless, they can be of practical use when implementing SOS servers and clients. For example, to decide which parts of the schemas to support, to suggest how to encode large datasets of observations, to know where to look for the sensors' location, just to mention some. As future work we are trying to use all of this information to build customized SOS servers and clients that allow large amounts of data to be handled efficiently.